# Unipolar transport in bilayer graphene controlled by multiple *p-n* interfaces

Hisao Miyazaki,[1,2] Song-Lin Li,[1] Shu Nakaharai,[3] Kazuhito Tsukagoshi[1,2,a]

[1] International Center for Materials Nanoarchitectonics (WPI-MANA), NIMS, Namiki, Tsukuba 305-0047, Japan

[2] CREST, JST, Honcho, Kawaguchi 332-0012, Japan

[3] GNC, AIST, Onogawa, Tsukuba, 305-8569, Japan

**Abstract: Unipolar transport is demonstrated in a bilayer graphene with a series of *p-n* junctions and is controlled by electrostatic biasing by a comb-shaped top gate. The OFF state is induced by multiple barriers in the *p-n* junctions, where the band gap is generated by applying a perpendicular electric field to the bilayer graphene, and the ON state is induced by the *p-p* or *n-n* configurations of the junctions. As the number of the junction increases, current suppression in the OFF state is pronounced. The multiple *p-n* junctions also realize the saturation of the**

---

[a] To whom correspondence should be addressed; E-mail: TSUKAGOSHI.Kazuhito@nims.go.jp.

**drain current under relatively high source-drain voltages.**

Graphene is a potential high-mobility channel for field effect transistors (FETs)[1-3]. However, the absence of a band gap has been one of the most serious issues in the application of graphene to ultra-fast logic circuits because it causes a serious leakage current in the OFF state. To overcome this issue, graphene nanoribbon (GNR)[4-9] and bilayer graphene (BLG)[10-17] have been developed to improve the OFF properties by increasing the band-gap energy. Although the enlargement of the band gap is possible, further improvement to realize the complementary metal oxide semiconductor (CMOS)-compatible circuits in graphene with low enough leakage currents is required. In a CMOS circuit[18,19], the ambipolar nature of graphene makes its operation unstable against the threshold voltage variation[20], which is due to a small gate voltage region for the OFF state. Therefore, a wide margin of the OFF state is desired for the stable operation of the logic circuits in graphene and for a large band gap. A possible solution for this demand is to extend the OFF-state span by suppressing one of the ambipolar branches of transport of positive (*p*) or negative (*n*) carriers to produce a unipolar behavior.

In this paper, we propose and demonstrate a unipolar-like FET based on bilayer graphene. The proposed structure has a comb-shaped top gate, which has multiple rectangular gates in series on the bilayer graphene channel. The local control of the carrier density and polarity by the top gate, which is accompanied by the global control of the back gate, determines the junction configurations as *p-n*, *p-p*, *n-p* or *n-n*[21-25]. When a large gate bias is applied to the BLG, a band gap is generated at the Dirac point, which causes a highly resistive region in the *p-n* junctions. As a result, the drain current in the *p-n* or *n-p* configuration is made smaller by the multiple *p-n* junctions than that in the *p-p* or *n-n* configuration. Because a top-gate sweep switches between the *p-n* (*n-p*) and *p-p* (*n-n*) configurations, a unipolar-like behavior, i.e., an asymmetric ambipolar behavior, is realized. The saturation behavior of the drain current at a relatively high source-drain bias is also discussed.

Flakes of BLG were prepared by micromechanical cleavage[1] of Kish graphite on a highly doped Si substrate with a 285-nm $SiO_2$ layer on top. The source and the drain electrodes (Au (30 nm) on Ti (5 nm)) were formed with a common channel length $L_{ch}$ of 5.0 μm, followed by the patterning of the BLG channel using oxygen plasma with a uniform channel width $W_{ch}$ (=0.5-1.0 μm). Then, a top gate electrode (50 nm Al) was directly deposited on the BLG. The top gate dielectric was formed using the natural oxidization of Al at the Al/BLG interface[26-28]. The top gate has a comb-shaped structure (Fig. 1(a)) so that the carrier density and polarity are spatially modulated repeatedly. The width of each comb finger is designed to be $L_g/N$, where $L_g$ is the total length of the top-gated region, and $N$ is the number of comb fingers ($N$ =1, 4, 6, 8); in this configuration, every device in the present experiment has a fixed total length of $L_g$=1.7 μm.

The setup for our measurement is illustrated in Fig. 1(b). A constant dc drain bias $V_{ds}$ (= 10 mV) was applied from the drain, while the source was grounded, and the drain current, $I_{ds}$, was measured by a simple two-terminal method. The top-gate bias, $V_{tg}$, and the back-gate bias, $V_{bg}$, were applied independently so that the carrier density and polarity of both the top-gated and the bare graphene regions could be controlled independently. The $V_{tg}$ and $V_{bg}$ dependences of the conductance between the source and the drain, $G_{ds}$, were found to be similar to previously reported values[11,15-17], except for the effects of multiple p-n junctions that we discuss in this letter. The samples were installed in a cryostat, and electrical characterization was performed at 80 K.

Figure 1(c) represents the $V_{tg}$-$V_{bg}$ mapping of $G_{ds}$ and some schematics of the carrier density profiles along the channel. This $V_{tg}$-$V_{bg}$ mapping is divided into four characteristic regions, where the top-gated and the bare regions have different carrier polarities, and their boundaries (thick broken lines) correspond to the charge neutrality points (CNPs) of the bare (horizontal line) and the top-gated (inclined line) regions of the BLG. The carrier density of each region is expressed by $n_b(V_{bg}) = C_{bg}(V_{bg} - V_{bg}^0)/e$ and $n_t(V_{tg}, V_{bg}) = C_{tg}(V_{tg} - V_{tg}^0)/e + n_b(V_{bg})$ for the bare and the top-gated regions, respectively, where $C_{bg}$ ($C_{tg}$) is the back- (top-) gate capacitance per unit area and $V_{bg}^0$ ($V_{tg}^0$) is the CNP shift for the back (top) gate[21] as indicated by the cross point of the two CNP lines. Here, positive and negative values of $n_t$ represent the density of electrons and holes, respectively. The conditions for the charge neutrality in the top-gated region is given by $n_t = 0$, which determines the top-gate bias for the charge neutrality of the top-gated region as a function of the back-gate bias, $V_{tg}^{CN} = -(C_{bg}/C_{tg})(V_{bg} - V_{bg}^0) + V_{tg}^0$. Note that the charge neutrality condition for the bare region is independent of the top-gate bias, $V_{bg}^{CN} = V_{bg}^0$. By taking the slope of the CNP line for $V_{tg}^{CN}$ in the $V_{tg}$-$V_{bg}$ mapping, $C_{tg}$ is obtained as $C_{tg} = C_{bg}/0.013 = \kappa_{SiO2}\varepsilon_0/0.013 d_{bg} = 0.93$ µF/cm², where $\kappa_{SiO2} = 3.9$ is the relative permittivity of $SiO_2$, $\varepsilon_0$ is the vacuum permittivity, and $d_{bg} = 285$ nm is the thickness of the back-gate dielectric. This value corresponds to an $SiO_2$-equivalent thickness of the top-gate dielectric, $d_{tg} = 0.013 d_{bg} = 3.7$ nm.

In Fig. 1(c), $G_{ds}$ of the *p-n* configuration is strongly suppressed compared with the cases of *p-p* and *n-n*. Such asymmetry has not been found in the case of single-top-gate BLG devices[15]. The cause of this effect is considered to be the multiple *p-n* junctions that are generated by the multiple top gates. In fact, in the case of the *p-n* configurations, the resistive *p-n* junctions are induced by the field-effect-induced band gap, which is more evident in a higher back-gate bias reaching about a threefold difference: 4.7 kΩ for *p-p* and 13.5 kΩ for *p-n* at a fixed carrier density of $|n_\mathrm{b}|=|n_\mathrm{t}|=6.4\times10^{12}$ cm$^{-2}$ at $V_\mathrm{bg}=-80$ V. This large difference cannot be explained by the difference in the carrier mobility because the difference between the electron mobility ($\mu_n=1700$ cm$^2$/Vs) and the hole mobility ($\mu_p=1900$ cm$^2$/Vs) is only 11%[29]. The difference of the resistance (8.8 kΩ) for 2*N* (=16) *p-n* junctions also implies that the resistance per junction is 0.6 kΩ, which is on the same order as the tunnel *p-n* junction (approximately 1 kΩ) in a BLG with a similar channel width[30]. Therefore, the asymmetry in the conductance, i.e., the unipolar-like behavior, is considered to be induced by the multiple *p-n* junctions with an electric-filed-induced band gap.

The effect of an electric-field-induced band gap on the unipolar behavior is analyzed in Fig. 2. Some slices of the $V_\mathrm{tg}$-$V_\mathrm{bg}$ mapping in Fig. 1(c) at a fixed $V_\mathrm{bg}$ exhibit asymmetric characteristics are shown in Fig. 2(a), where $V_\mathrm{tg}$ at the $G_\mathrm{ds}$ minimum corresponds to the CNP, $V_\mathrm{tg}^\mathrm{CN}$. To characterize the asymmetry, we consider a resistance difference, Δ*R*, between the left and the right branches[21,31], as illustrated in the inset of Fig. 2(a). The difference is extracted from two points that are equally separated from the CNP by $\Delta V_{tg} = |V_{tg} - V_{tg}^{CN}|$. When the bare regions are $p$-type, $\Delta R$ represents the resistance difference between the $p$-$n$ ($R_{p\text{-}n}$) and the $p$-$p$ ($R_{p\text{-}p}$) configurations, i.e., $\Delta R = R_{p\text{-}n} - R_{p\text{-}p}$. Similarly, $\Delta R = R_{n\text{-}n} - R_{n\text{-}p}$ for the $n$-type bare regions. Figure 2(b) shows $\Delta R$ as a function of $\Delta V_{tg}$. The value $\Delta R$ becomes larger as the absolute value of the back-gate bias increases ($|V_{bg}| \to 80\text{ V}$). This correspondence reflects the enhancement of the $p$-$n$ junction resistance by the electric-field-induced band gap, which is more dominant than the effect of the carrier-density increase in the bare regions caused by the increase in the back-gate bias. However, the decrease in $\Delta R$ for large $\Delta V_{tg}$ is not caused by shrinking of the band gap because the band gap becomes larger when $\Delta V_{tg}$ increases in the case of the $p$-$n$ configuration. Rather, the decrease in $\Delta R$ could be attributed to the carrier-density increase in the top-gated regions and/or the shortening of the tunnel barrier, which is caused by the steeper potential slope at the $p$-$n$ junctions[32]. When $V_{bg}$>approximately 0 V, $\Delta R$ becomes negative as shown in the inset of Fig. 2(b), reflecting the inverted carrier polarity in the bare regions. The reason is that $\Delta R = R_{n\text{-}n} - R_{n\text{-}p}$ for the $n$-type bare regions, and the $p$-$n$ junction resistance makes $R_{n\text{-}p}$ greater than $R_{n\text{-}n}$.

The contribution of the number of junctions ($N$) to the unipolar characteristics is presented in Fig. 3(a). Here, the $N$ dependence of the normalized conductance, $G_{ds}/W_{ch}\mu_h$, is plotted as a function of $n_t$ at $V_{bg} - V_{bg}^0 = -75$ V, while the total length of the top-gated region is fixed as $L_g$ for all devices with different $N$ values. In Fig. 3(a), the unipolar behavior is evident with larger $N$. This $N$-dependence only appears when the band gap is generated by a large $|V_{bg} - V_{bg}^0|$. In Fig. 3(b), the ratio of the maximum slope of the $G_{ds}$-$V_{tg}$ curve for the $p$-$p$ branch ($n_t < 0$) to that for the $p$-$n$ branch ($n_t > 0$) is plotted as a function of $N$ at different back-gate biases, $V_{bg} - V_{bg}^0$. There is no $N$-dependence for the zero back-gate bias, which reflects a small junction resistance due to the small band gap. When the back-gate bias is larger, the maximum slope ratio increases with $N$, reflecting a larger junction resistance due to the larger band gap.

Another advantage of our device structure is the saturation behavior of the source-drain current $I_{ds}$ at a large source-drain voltage $V_{ds}$ with larger $N$ values. The increase of $I_{ds}$ has been reported to be suppressed in the ON state when the local charge neutrality point is at the edge of a top-gated region by $V_{ds} \sim |V_{tg} - V_{tg}^{CN}|$ [33,34]. The current saturation can be induced more effectively in our device with a larger $N$ than that in a single top-gate device because the local charge neutral takes place at each top-gate-finger edge successively as the $V_{ds}$ increases (Fig. 4(d)). Figure 4(c) shows the comparison between the $V_{ds}$-$I_{ds}$ curves of the devices with single (Fig. 4(a)) top gates and those with multiple (Fig. 4(b)) top gates at $V_{bg} - V_{bg}^0 = -80$ V and $V_{tg} - V_{tg}^0 = -0.8$

V. The saturation behavior in the device with multiple top gates is more evident than that in the single-gated device.

In summary, we have demonstrated the unipolar operations in FETs based on a bilayer graphene with a comb-shaped top gate. The unipolar behavior was realized by the suppressed conductance in the OFF state, which was induced by a series of resistive *p-n* junctions, whereas in the ON state, the junctions were in the *p-p* or *n-n* configurations without high resistive regions. The high resistance in the *p-n* junction was shown to be generated by the electric-field-induced band gap by extracting the gate bias dependence of the asymmetric conductance. The effect of the multiple gates was also shown by the junction number dependence of the conductance asymmetry. Finally, the saturation characteristics of the source-drain current in a large source-drain bias were presented as an additional advantage of the proposed device concept.

**ACKNOWLEDGMENTS**

This work was supported in part by KAKENHI (No. 21241038) from the MEXT of Japan and by the FIRST Program from the JSPS. We thank Covalent Materials Corporation for providing Kish graphite as the source material of graphene.

Figure 1: (a) False color AFM image of a device with finger number $N=8$, and (b) a schematic diagram of the device. (c) Contour plot of the source-drain conductance $G_{ds}$ as a function of $V_{bg}$ and $V_{tg}$ for an $N=8$ device measured at $T=80$ K. The white lines indicate the CNPs for the bare (vertical line) and the top-gated (diagonal line) regions. Separated by the lines, the $V_{bg}$-$V_{tg}$ space is categorized as the *p-n*, *p-p*, *n-n*, and *n-p* configurations. The schematically shown charge distributions are related to the four configurations.

Figure 2: (a) $G_{ds}$ as a function of $V_{tg}$ for $N=8$ at $V_{bg}=0$, −20, −40, −60, and −80 V, as extracted from Fig. 1(c). The inset shows the definition of $\Delta R$, which is the resistance difference between the *p-n* and the *p-p* configurations for the *p*-type bare regions. (b) $\Delta R$ as a function of $\Delta V_{tg}$ for the data in (a). (c) Color plot of $\Delta R$ as a function of $V_{bg}$ and $\Delta V_{tg}$. The top markers indicate the corresponding $V_{bg}$ in (a) and (b). In the gray-colored region, $\Delta R$ is undefined.

Figure 3: (a) Source-drain conductance as a function of the top-gate voltage for $N=1$, 4, 6, and 8 at $V_{bg} - V_{bg}^0 = -75$ V ($n_b = -6.0 \times 10^{12}$ cm$^{-2}$). The conductance is normalized by $W_{ch}\mu_h$. The top-gate voltage is normalized by $1/C_{tg}$ and is shifted so that the *p-p*

branches align with each other. (b) Maximum $|dG_{ds}/dV_{tg}|$ ratio of the *p-p* branch ($V_{tg} < V_{tg}^{CN}$) to the *p-n* branch ($V_{tg} > V_{tg}^{CN}$) as a function of $N$ for $V_{bg} - V_{bg}^{0} = 0, -25, -50,$ and $-75$ V. All data were acquired at $T = 80$ K.

Figure 4: Output ($I_{ds}$-$V_{ds}$) characteristics ranging from $V_{tg} - V_{tg}^{CN} = 0$ to $-2.0$ V at $-0.4$ V step for $N=1$ (a) and $N = 6$ (b) samples, which are fabricated on the same BLG film at $V_{bg} - V_{bg}^{0} = -80$ V and $T=80$ K. (c) Identical data extracted from (a) and (b) for $V_{tg} - V_{tg}^{CN} = -0.8$ V. (d) Schematic diagram of the spatially modulated charge density in the unbiased ($V_{ds}=0$, left) and the biased ($V_{ds}<0$, right) cases.

Fig. 1

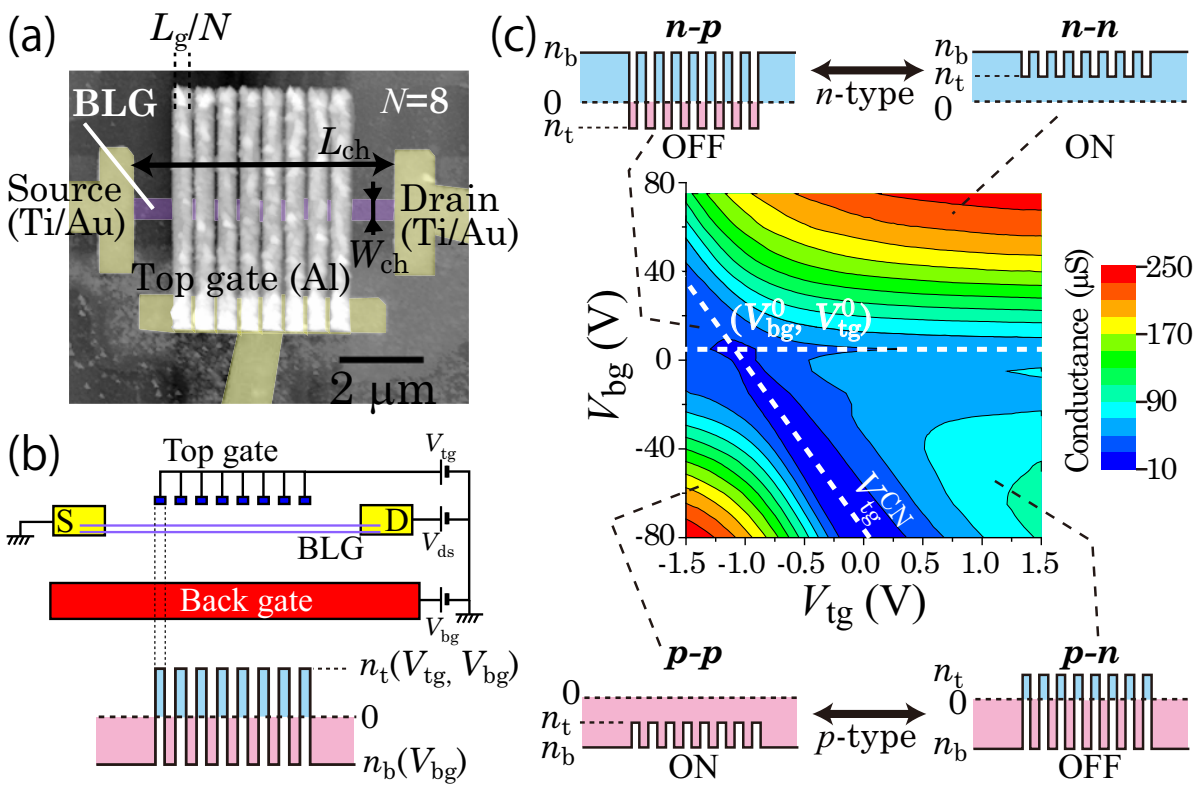

Fig. 2

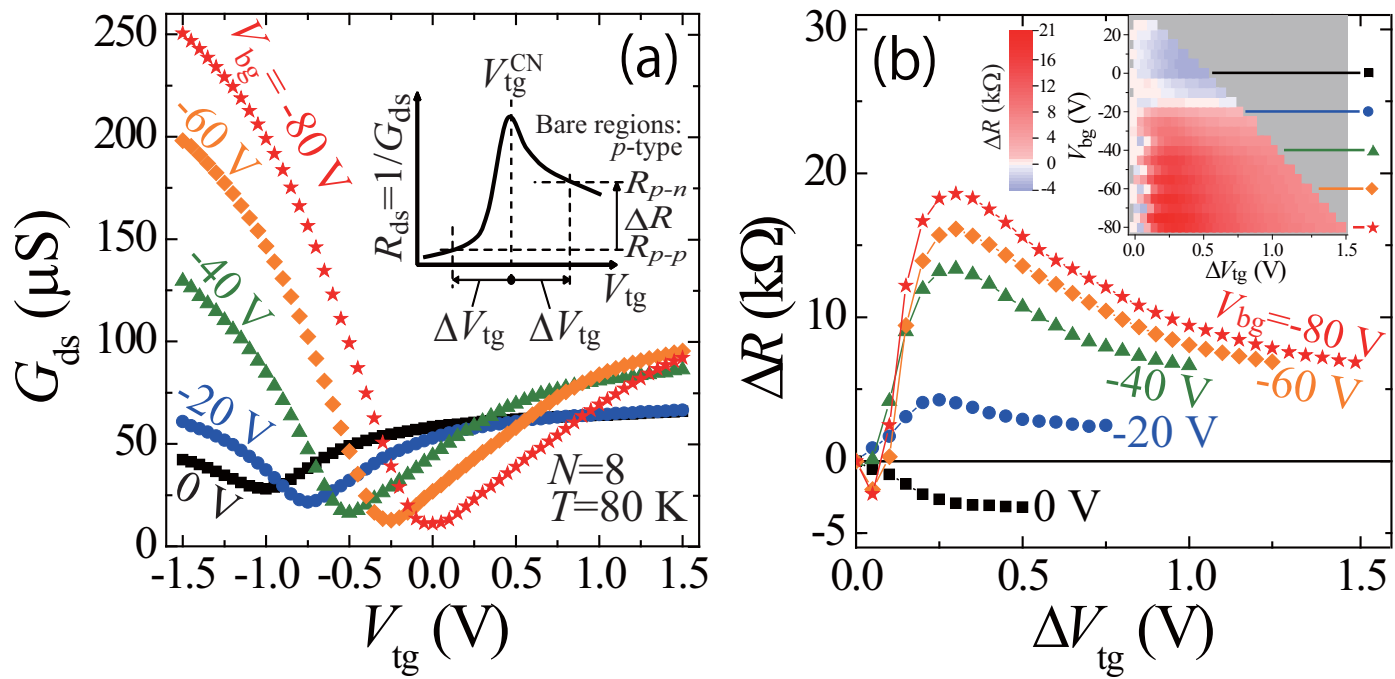

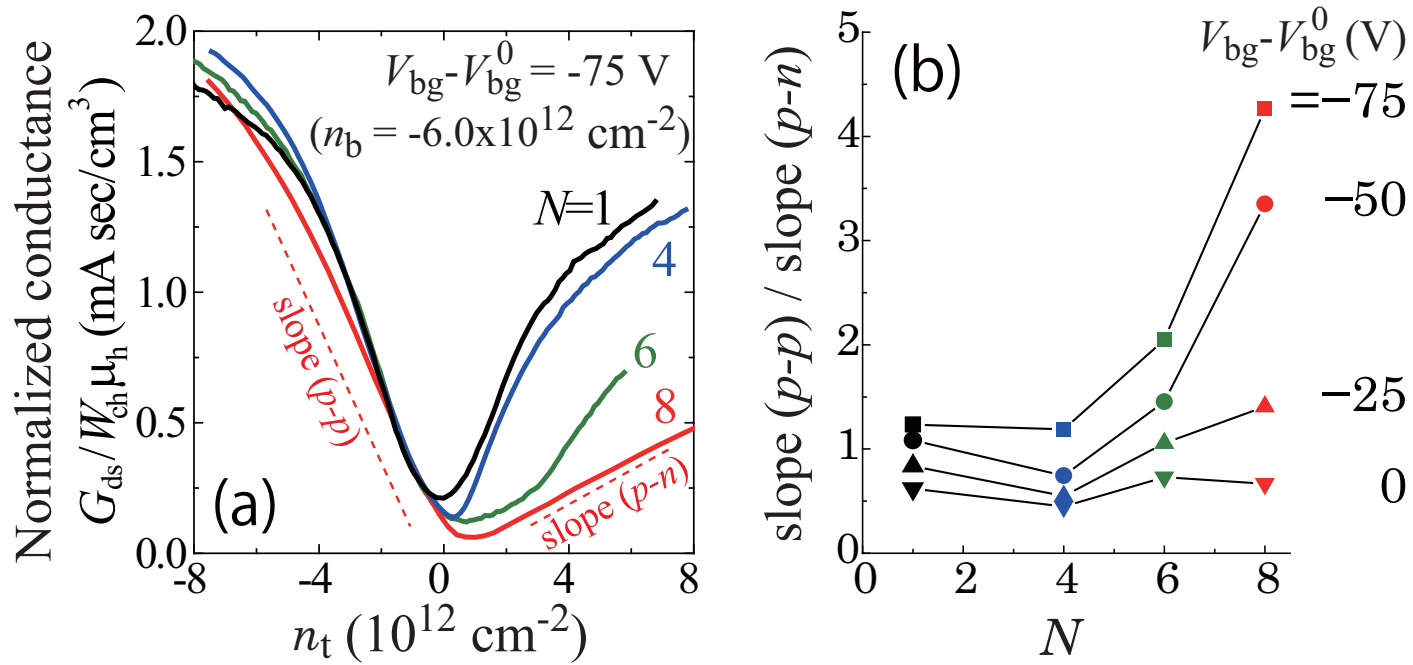

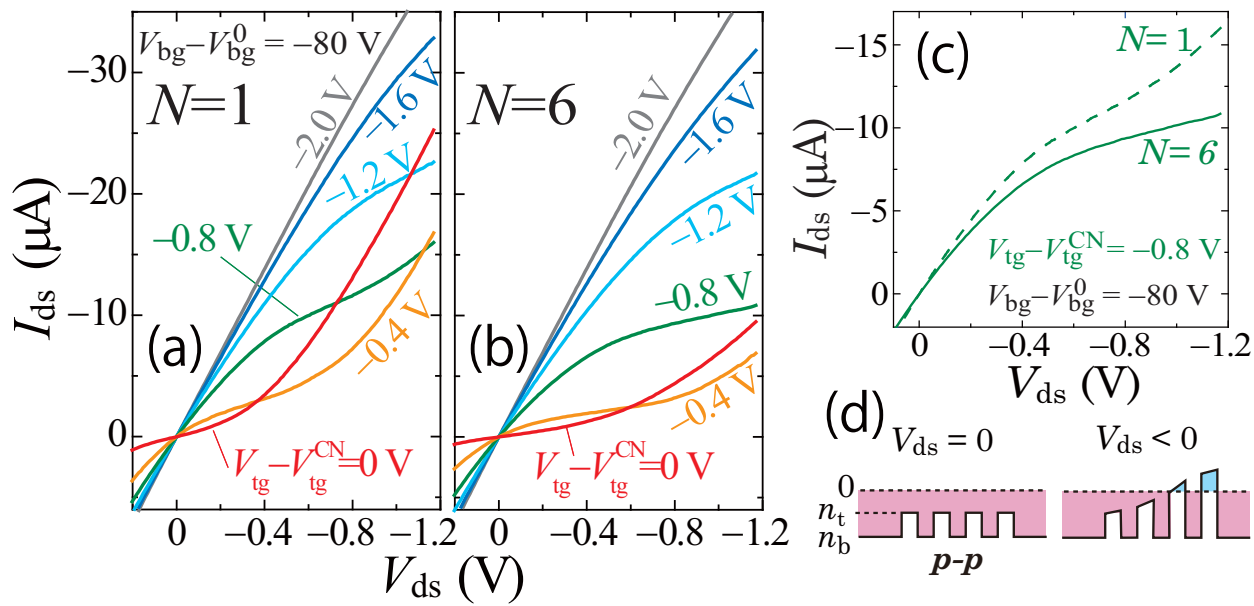

Fig. 4